\documentclass[aps,reprint,article,prx]{revtex4-1}
\usepackage{amsmath}
\usepackage{dcolumn}
\usepackage{graphicx}
\usepackage{float}
\usepackage{color}
\usepackage{ulem}
\begin{document}
\title{Vanishing Vortex Creep at the Transition from Ordered to Disordered Vortex Phases in Ba$_{0.64}$K$_{0.36}$Fe$_2$As$_2$}

\author{Yu-Hao Liu,$^1$ Wei Xie,$^{1,2}$ and Hai-Hu Wen$^{1,\dag}$}

\address{$^{1}$National Laboratory of Solid State Microstructures and Department of Physics, Collaborative Innovation Center of Advanced Microstructures, Nanjing University, Nanjing 210093, China \\ $^{2}$Western Superconducting Technologies Company Ltd., Xi’an 710018, China}

\begin{abstract}
By measuring the dynamical and conventional magnetization relaxation of the Ba$_{0.64}$K$_{0.36}$Fe$_2$As$_2$ single crystals, we found strong second peak effect on the magnetization hysteresis loops. It is found that there is a kink of magnetization at a field between the valley and maximum magnetization. Interestingly, the magnetization relaxation rate has a deep minimum at the field with the kink, indicating a diminished vortex creep. The relaxation rate at this field is clearly smaller than the so-called universal lower limit of the relaxation rate characterized by \(S_0\approx Gi^{1/2}(T/T_\mathrm{c})\). This diminished vortex creep is associated with the origin of the SMP effect and attributed to the strongly hindered flux motion when experiencing the transition from the quasi-ordered to disordered vortex phases.
\end{abstract}

\maketitle
\section{Introduction}
Since the discovery of superconductivity in LaFeAsO$_{1-x}$F$_x$ \cite{ref1}, the iron-based superconductors have attracted considerable attention due to their unique properties and potential applications. Among the iron-based superconductors, the optimally doped Ba$_{1-x}$K$_x$Fe$_2$As$_2$ single crystal has a high superconducting transition temperature ($T_\mathrm{c}$) \cite{ref2,ref3}, large critical current density ($J_\mathrm{s}$) \cite{ref4,ref5,ref6}, a small anisotropy ratio ($\gamma$) \cite{ref7,ref8,ref9}, and samples are easy to be fabricated \cite{ref2,ref3}.A lot of studies on vortex dynamics were carried out on this type of samples \cite{ref4,ref5,ref6,ref10,ref11,ref12}. One of the most interesting phenomena observed in the Ba$_{1-x}$K$_x$Fe$_2$As$_2$ single crystals is the second magnetization peak (SMP) effect, in which the width of magnetic hysteresis loops (MHLs) will increase with increasing magneticand a second peak emerges far away from the zero field. This is of great significance for the high-field application of the iron-based superconductors.

The SMP can be divided into three types according to the different shapes of MHLs. The first type occurs mainly in conventional superconductors, such as Nb \cite{ref13}, Nb$_3$Sn \cite{ref14}, CeRu$_2$ \cite{ref15}, 2H-NbSe$_2$ \cite{ref15,ref16}, and heavy fermion compounds \cite{ref17}. The origin of this type is generally explained as the thermal melting of vortex lines and the enhancement of flux pinning \cite{ref18,ref19}, and the peak position is near the upper critical field $H_\mathrm{c2}$.The second type occurs mainly in highly anisotropic cuprate superconductors, such as Bi$_2$Sr$_2$CaCu$_2$O$_{8+\delta}$ (Bi2212) \cite{ref20}, Bi$_2$Sr$_2$CuO$_{6+\delta}$ (Bi2201) \cite{ref21}, and the peak in MHLs locates in a weak magnetic field and is temperature independent.The third type mainly occurs in YBa$_2$Cu$_3$O$_y$ (YBCO) \cite{ref22}, Ba$_{1-x}$K$_x$BiO$_3$ (BKBO) \cite{ref23}, and some iron-based superconductors \cite{ref4,ref5,ref6,ref24}, which is often referred to as the fishtail effect due to its shape, and the second peak position is strongly temperature dependent.Several theories have been proposed to explain the origin of the SMP effect, such as a crossover from elastic to plastic pinning as the field increases \cite{ref25,ref26,ref27}, structural phase transition of vortex lattices \cite{ref28,ref29}, and order-to-disorder vortex phase transition \cite{ref22,ref30,ref31,ref32}, etc.

In YBa$_2$Cu$_3$O$_{7-\delta}$ (YBCO) crystals, a sharp kink has been observed between the valley (\(H_{\text{dip}}\)) and the second peak (\(H_\mathrm{p}\)) of MHLs, and the field was named \(H_{\text{kink}}\) where the magnetic-field derivative $dm/dH$ is the largest \cite{ref33, ref34, ref35}. This phenomenon was interpreted as an order-disorder vortex phase transition and regarded to have a strong correlation with the origin of the SMP effect.Similarly, the Bragg peak has been observed in Ba$_{0.64}$K$_{0.36}$Fe$_2$As$_2$ single crystal by small-angle neutron scattering (SANS) measurement, which means a long-range ordered lattice exists at low field. As the magnetic field increases, the sharp Bragg peak crosses over to a diffraction ring, which was interpreted as an order-disorder transition concerning the SMP effect in MHLs \cite{ref30}.Meanwhile, the vortex structure with short-range hexagonal order was observed in optimally doped Ba$_{0.6}$K$_{0.4}$Fe$_2$As$_2$, such as the scanning tunneling microscopy (STM) \cite{ref36} at high field (\(H = 9\, \text{T}\)) and magnetic force microscopy (MFM) \cite{ref37} at low field (\(H < 100\, \text{Oe}\)).

In this study, we carry out systematic investigations on the magnetization relaxation rate of high-quality Ba$_{0.64}$K$_{0.36}$Fe$_2$As$_2$ single crystal. In the MHLs, a pronounced SMP effect and the kink can be observed. In the studies on magnetization relaxation, a minimum value of magnetization relaxation rate can be observed at \(H_{\text{kink}}\). The relaxation rate at the kink field is extremely small, indicating a very weak vortex motion. This observation provides evidence suggesting a potential association between the order-disorder transition and the SMP effect of iron-based superconductors.

\section{Experimental Detalls}
The Ba$_{0.64}$K$_{0.36}$Fe$_2$As$_2$ single crystals were grown by the self-flux method \cite{ref2}, with dimensions $1.6\, \text{mm} \times 1.4\, \text{mm} \times 0.14\, \text{mm}$. The composition of the sample was characterized using scanning electron microscopy (SEM) conducted on Phenom ProX (Phenom). The DC magnetization measurements were carried out with a SQUID-VSM-7T (Quantum Design). The magnetic field \(H\) was applied parallel to the \(c\) axis of the single crystal.In the dynamical magnetization relaxation measurements \cite{ref38}, the magnetic hysteresis loops (MHLs) were measured in magnetic fields up to \(7\, \text{T}\) with different sweeping rates (${dH}/{dt}$ = 200\,\text{Oe}/50\,\text{Oe}). In the conventional magnetization relaxation measurements \cite{ref39}, the sample was zero-field cooled down from above \(T_\mathrm{c}\) to the desired temperature, and then the magnetic field was quickly increased to a certain value. The magnetization measurements started immediately after the certain fields were applied, and the measuring time is \(3\, \text{hours}\).
\section{RESULTS AND DISCUSSION}
\begin{figure}[htbp]
\centering
\includegraphics[width=8.6cm]{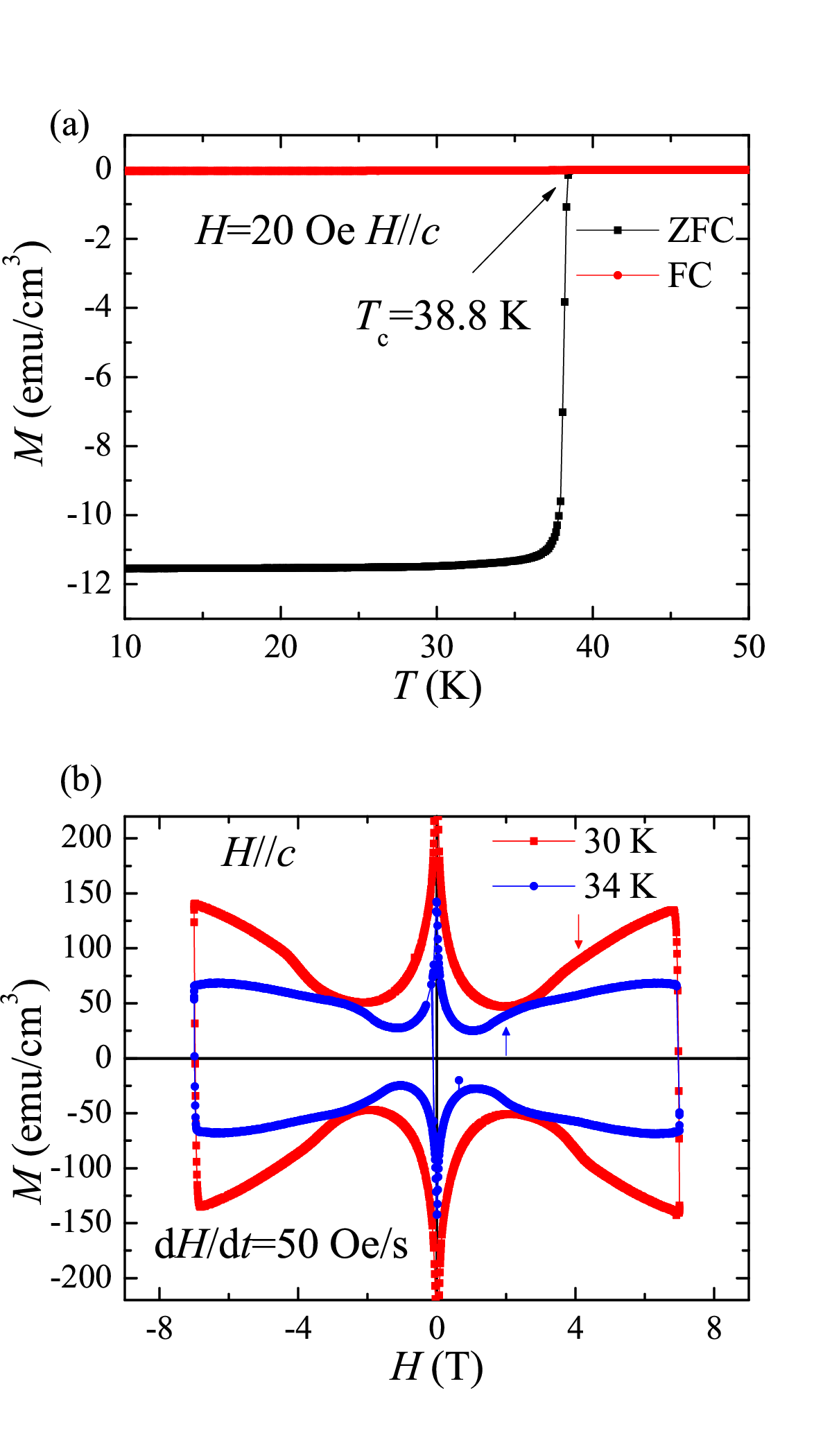}
\caption{Temperature dependence of magnetization and typical magnetization hysteresis loops (MHLs) of Ba$_{0.64}$K$_{0.36}$Fe$_2$As$_2$ single crystal. (a) The superconducting transition curve obtained with zero-field cooled and field cooled modes at $H=20\text{Oe}$ with $H//c$. (b)The isothermal MHLs of the sample with $dH/dt=50\text{Oe/s}$ at 32K (red) or 34K (blue) . The arrows indicate the kink positions on MHLs.
} \label{fig1}
\end{figure}

Fig.~\ref{fig1}(a) shows the temperature dependence of magnetization of the Ba$_{0.64}$K$_{0.36}$Fe$_2$As$_2$ single crystal with the zero-field cooled (ZFC) and field cooled (FC) modes. The magnetic field is $20\, \text{Oe}$ and parallel to the $c$-axis of the sample. The sharp transition indicates that the sample is of high quality. The obtained superconducting transition temperature is $38.8\, \text{K}$, which is determined as the intersection of the magnetization in ZFC and FC modes in Fig.~\ref{fig1}(a).The magnetic hysteresis loops (MHLs) for the Ba$_{0.64}$K$_{0.36}$Fe$_2$As$_2$ single crystal were measured at different temperatures. The results measured at $30\, \text{K}$ or $34\, \text{K}$ with ${dH}/{dt}$ = 200\,\text{Oe}/50\,\text{Oe} are shown in Fig.~\ref{fig1}(b). The MHLs exhibit a symmetric shape as a result of strong bulk pinning. The values of $H_{\text{kink}}$ at $30\, \text{K}$ and $34\, \text{K}$ are $4.1\, \text{T}$ and $2\, \text{T}$, respectively. The $H_{\text{kink}}$ of the Ba$_{0.64}$K$_{0.36}$Fe$_2$As$_2$ in Fig.~\ref{fig1}(b) decreases gradually with increasing temperature, which is different from YBCO \cite{ref35}, and this may be caused by different pinning mechanisms of the two superconductors.

\begin{figure}[htbp]
\centering
\includegraphics[width=8.6cm]{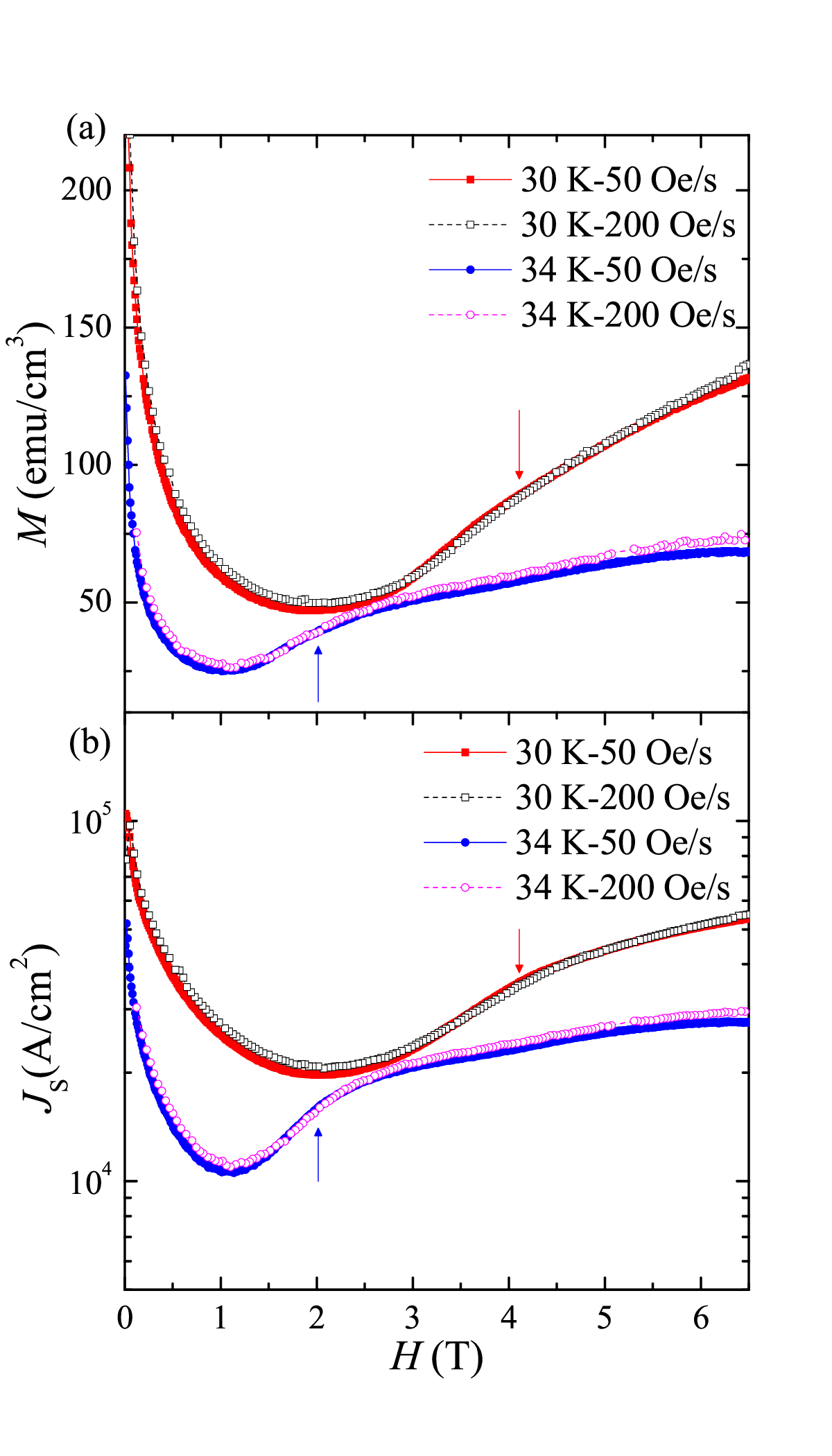}
\caption{Correlations of (a) MHLs and (b) $J_\mathrm{s}$ versus $H$ for Ba$_{0.64}$K$_{0.36}$Fe$_2$As$_2$ single crystal at different temperatures with $dH/dt = 200 \text{Oe/s}$ (open) and $50 \text{Oe/s}$ (solid). Arrows indicate the closest or intersecting position of the curves with different field sweeping rates.
} \label{fig2}
\end{figure}

Shown in Fig.~\ref{fig2}(a) are the magnetic hysteresis loops (MHLs) measured in the field ascending processes with different field sweeping rates. At each temperature, an anomalous intersection between $M(H)$ curves with different field sweeping rates is observed. To eliminate the influence of equilibrium magnetization, we calculated the superconducting current density by the Bean critical state model \(J_s\)=20\(\Delta \)\(M\)/\(w\)(1-\(w\)/3\(l\)), where \(\Delta \)\(M\)= \(M_+\)-\(M_-\)), \(M_+\)(\(M_-\)) is the magnetization associated with descending (ascending) field, \(l\) is the length, and \(w\) is the width of the single crystal (\(l > w\)) \cite{ref40}. The results are shown in Fig.~\ref{fig2}(b).As one can see, double crossings of the isothermal MHL curves occur in the near region of the kink at \(H_{\text{kink}}\). It shows that the magnetization width \(\Delta \)\(M\) is even getting narrower measured with a high sweeping rate than that with a lower sweeping rate. This is abnormal, indicating a “negative” relaxation rate in the field sweeping process. In the field sweeping process, it is known that, if \(dM/dt\) is much smaller than \( E \propto dB/dt\) , thus combining with the thermally activated flux creep model
\begin{equation}\label{equ1}
E = v_0Bexp[-\frac{U(T,j)}{k_\mathrm{B}T}]
\end{equation}
and the Kim-Anderson model, \(U(T,j)\)=\(U_\mathrm{c}(T)\)\((1-j/j_\mathrm{c}(T))\), one can get the relation
\begin{equation}\label{equ2}
j=j_\mathrm{c}(T)[1-\frac{k_\mathrm{B}T}{U_\mathrm{c}(T)} \ln(\frac{v_0B}{E})]
\end{equation}

Where \(U(T, B)\) is the activation energy, \(v_0\) is the attempting hopping velocity. One can see that a larger field sweeping rate should correspond to a larger \(E\), and a larger transient current density \(j\). And according to the Bean critical state model, the width of the magnetic hysteresis loop (MHL) should be larger. This clearly indicates that the MHLs measured with a higher field sweeping rate should always be wider than that with a lower sweeping rate. Thus, the crossings of the MHLs measured with different sweeping rates shown in Fig.~\ref{fig2} are abnormal. For a thin disk, \(dM/dt << dB/dt\), a good approximation can be established \cite{ref41,ref42}
\begin{equation}\label{equ3}
U(T,B)=k_\mathrm{B}T ln(\frac{2v_0B}{w(dB/dt)}).
\end{equation}
Here \(w\) is the width of the sample. Eq.~\ref{equ3} indicates that a faster sweeping rate corresponds to a smaller activation energy, thus expecting a larger \(j\) value and a wider magnetic hysteresis loop (MHL), again showing that the intersection of MHLs with different field sweeping rates is unreasonable.Taking into account the non-uniform distribution of the magnetic field in the sample, we believe this may be a fake observation, namely a smaller width of MHL corresponds to a larger sweeping rate. But at least, it shows a negligible relaxation effect near the kink field. To comprehend the physics concerning the flux dynamics near the kink, we calculate the dynamical magnetization relaxation rate \(Q\), which is defined as
\begin{equation}\label{equ4}
Q=\frac{d ln(J_\mathrm{s})}{d ln(dB/dt)}=\frac{d ln(\Delta M)}{d ln(dB/dt)}.
\end{equation}
The parameter $Q$ is calculated by Eq.~\ref{equ4}, in which $J_\mathrm{s}$ is measured with two different field sweeping rates 50 and 200 Oe/s. The results of 30K, 32K, and 34K are shown in Fig.~\ref{fig4}(a). One can see negative values of $Q$ near the kink on the MHL.
\begin{figure}[htbp]
\centering
\includegraphics[width=8.6cm]{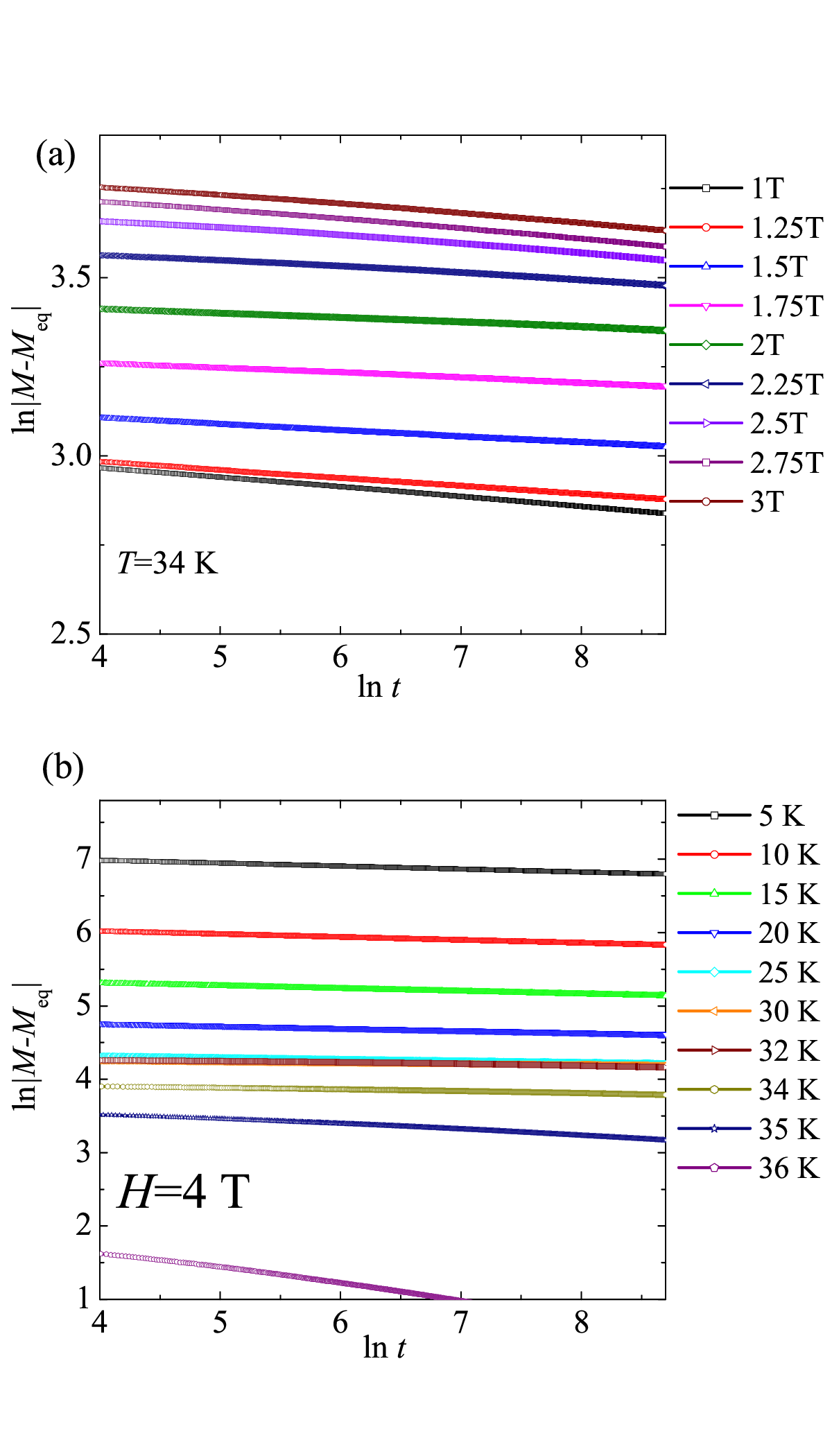}
\caption{Double-logarithmic plots of the time dependence of magnetization at (a) various fields at $T = 34 \text{K}$ and (b) different temperatures at $H=4 \text{T}$.
} \label{fig3}
\end{figure}

To better understand the physics concerning the vortex motion near the kink, we measured the magnetization relaxation and determined the relaxation rate $S$, namely the time dependence of magnetization is measured. The $S$ is determined through 
\begin{equation}\label{equ5}
S=-\frac{d ln(|M-M_\mathrm{eq}|)}{d lnt}
\end{equation}
where $M_\mathrm{eq}=(M_+ + M_-)/2$ is the equilibrium magnetization obtained from the MHLs at the same temperature. In order to investigate the field dependence of $S$, we measured the $M(t)$ at 30K, 32K, and 34K with different applied fields. The calculated results are shown in Fig.~\ref{fig4}(a) by the solid symbols. For each field and temperature, the measuring time was 10800s (3h). The double-logarithmic plots of $M(t)$ at 34K are shown in Fig.~\ref{fig3}(a). When the field increases to 2T, the slope decreases to a small value. This suggests a clear smaller conventional relaxation rates at 2T when the temperature is $T$ = 34K as shown in Fig.~\ref{fig4}(a). Similarly, we measured the $M(t)$ with a certain field at different temperatures. Three fields are chosen: 2\text{T}, 3T, 4T, and the results of 4T are shown in Fig.~\ref{fig3}(b). The slope of the curve decreases with increasing temperature below 30K, but above 30K the opposite trend is observed, indicating a smaller relaxation rate near the field of temperature where the kink of the MHL appears. Thus both the dynamical relaxation and conventional relaxation of magnetization all indicate a strongly suppressed relaxation rate near the kink, which we attribute to crossover of different vortex phases. 

In the single-vortex regime, the activation energy $U_\mathrm{sv}$  is equal to the pinning energy $E_\mathrm{pin}$  which can be expressed as $E_\mathrm{pin}=(\gamma \xi^2 L_\mathrm{c})^{1/2}$. Here $\gamma$ is the disorder parameter, $\xi$ is the coherence length and $L_\mathrm{c}$ is the collective pinning length. We can express $U_\mathrm{sv}=H_\mathrm{c}^2 \xi^4/L_\mathrm{c}$  via the elastic energy $U_\mathrm{sv}=U_\mathrm{e}=\epsilon_0 \xi^2/L_\mathrm{c}$ and $H_c=\Phi_0/2\sqrt{2}\pi \lambda \xi$, here $H_\mathrm{c}$ is the thermodynamic critical field, $\epsilon_0$ is an important energy scale which determines the self-energy of the vortex lines and can be given by $\epsilon_0=(\Phi_0/4\pi \lambda)^2$, $\lambda$ is the penetration depth and $\Phi_0=h/2e$ is the flux quantum. By introducing the Ginzburg number in the form of $Gi=(T_\mathrm{c}/H_\mathrm{c}^2\epsilon \xi^3)^2/2$ and a very common relation $(\epsilon \epsilon_0 \xi/T_\mathrm{c})^2=(1-T/T_\mathrm{c})/8Gi$, with $\epsilon$ is the anisotropy parameter, the activation energe $U_\mathrm{sv}$ can be rewritten as $U_\mathrm{sv}=T_\mathrm{c}\xi[(1-t)/Gi]^{1/2}/L_\mathrm{c}$, here $t$ is the reduced temperature $T/T_\mathrm{c}$\cite{ref43}. When the current $j$ reaches the critical current density $j_\mathrm{c}$, the pinning force $(\gamma L_\mathrm{c})^{1/2}$ is equal to the Lorentz force $j_\mathrm{c}\Phi_0L_\mathrm{c}$, then we can obtain the $\xi/L_\mathrm{c}=(j_\mathrm{c}/j_\mathrm{0})^{1/2}$, here $j_0$ is the depairing current density. Finally, the $U_\mathrm{sv}$ can be written as $T_\mathrm{c}[(1-t)/Gi]^{1/2}(j_\mathrm{c}/j_\mathrm{0})^{1/2}$\cite{ref19}. The effective pinning energy is typically defined in the form of $U=T/S$, so the $S=t[Gi/(1-t)]^{1/2}(j_0/j_\mathrm{c})^{1/2}$ is considered to be larger than $Gi^{1/2}(T/T_\mathrm{c})$ due to $j_0>j_\mathrm{c}$ and $t<1$\cite{ref19,ref44}. A noteworthy observation on curve $S(T,B)$ is that there is a clear minimum. Taking the related parameters of present superconductor, we found that $S(T,B)$ should increases linearly with temperature $T$. The comparison of $S(T)$ and $Gi^{1/2}(T/T_\mathrm{c})$ is shown in Fig.~\ref{fig4}(b). It is evident that the minimum relaxation rate is even lower than the lower limit of $Gi^{1/2}(T/T_\mathrm{c})$. We attribute this very low relaxation rate to the heavily damped vortex motion near the kink region, thus it is most likely related to a vortex phase transition. Considering the ordered vortex pattern observed in the low field region, we anticipate that the vortex structure undergoes a transition from a quasi-ordered (like Bragg glass state) to a more disordered state. It is the enhanced entanglement of vortices near the kink that produces the very small relaxation rate, as shown in Fig.~\ref{fig4}(a). It is found that once the second peak effect appears on the MHL curve, the local minimum of the relaxation rate emerges. This implies that once the magnetic field surpasses $H_\mathrm{dip}$, the vortex creep cannot be described by the single vortex pinning of collective pinning model anymore, and strong entanglement of vortices starts to occur. When the field is higher than $H_\mathrm{kink}$, the relaxation rate increases again for the system going to another phase, most likely in the vortex glass state.

Although $Q$ and $S$ are obtained through two different methods, we can see that they are almost equal to each other providing similar information of vortex dynamics \cite{ref42,ref45}. Fig.~\ref{fig4}(a) shows the field dependence of $Q$ and $S$. They both exhibit a minimum value at the same field indicated by the arrows. The position of the minimum coincides with the $H_\mathrm{kink}$ at all temperatures. Hence, it reveals that the kinks observed in MHLs and the minima observed in the relaxation rate curves are attributable to the same vortex structure transformations. In YBa$_2$Cu$_3$O$_{7-\delta}$ (YBCO) crystals, kinks were also observed on the magnetic hysteresis loop (MHL) curves and interpreted as an order-disorder vortex phases transition \cite{ref34}. At low temperatures and fields, the vortex structures often exhibit a quasi-ordered state, such as the Bragg glass. However, as the temperature increases or the magnetic field intensifies, a more disordered vortex phase is observed. There are different physical considerations, and thus expressions for the specific location of the phase transition, such as the magnetic field at which the vortex elastic energy \(E_{\text{el}}\) equals the pinning energy \(E_{\text{pin}}\) \cite{ref34,ref35} or the Bragg peak completely disappears \cite{ref30}. These descriptions, in essence, are consistent with each other. As the temperature or the field increases, the flux lines become entangled and disordered. Simultaneously, the relaxation rate of vortex creep decreases, leading to an increase in the critical current density. When the field approaches \(H_\mathrm{p}\), the system is still in the vortex glass state, and the relaxation rate increases or remains flat, but the critical current density becomes higher due to the enhanced entanglement of vortices. However, when the field is beyond the peak field of magnetization, the flux lines start to break, yielding some dislocations of the vortex, and the system enters the plastic motion regime. In this regime, the relaxation rate increases rapidly, and the critical current density drops down quickly.
\begin{figure}[htbp]
\centering
\includegraphics[width=8.6cm]{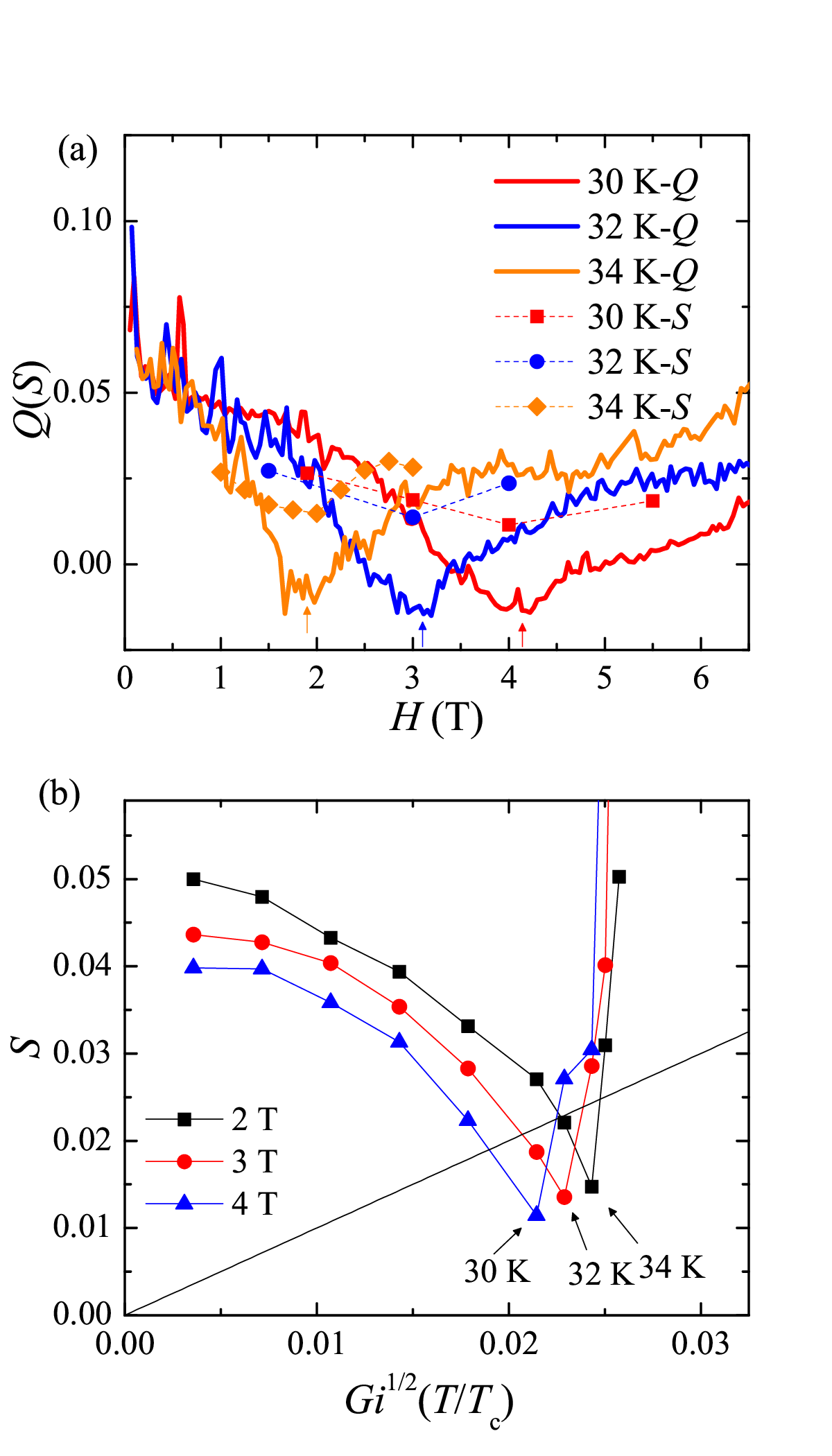}
\caption{(a) Field dependence of relaxation rate $Q$ (solid line) and $S$ (dashed line) at different magnetic temperatures. Arrows indicate the minimum values of $Q$ and $S$ plots. (b) The plots of $S$ versus $Gi^{1/2}(T/T_\mathrm{c})$ , which indicates a characteristic temperature that $S$ is minimum exists at a fixed magnetic field.
} \label{fig4}
\end{figure}

\begin{figure*}[htbp]
\centering
\includegraphics[width=17.2cm]{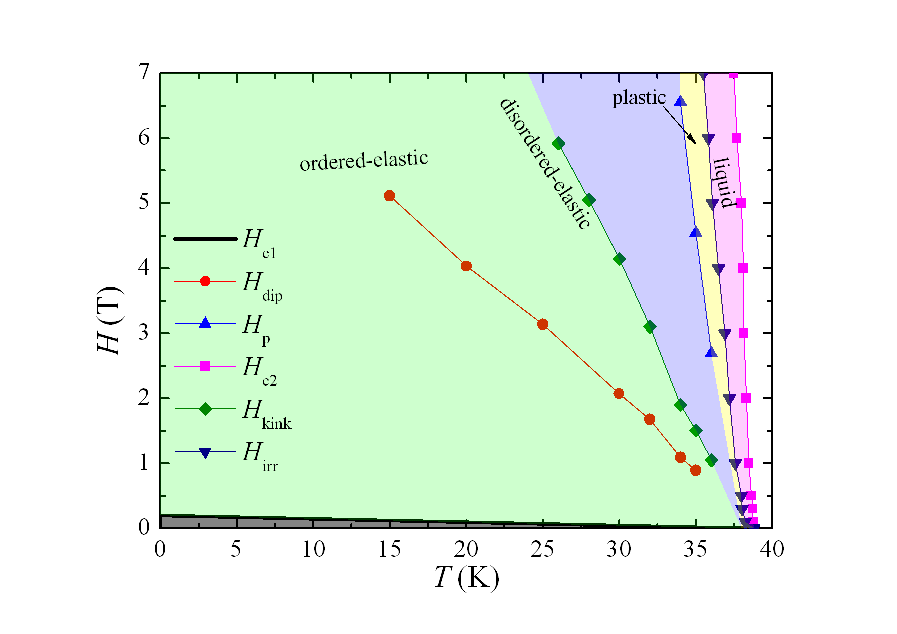}
\caption{Phase diagram of Ba$_{0.64}$K$_{0.36}$Fe$_2$As$_2$ single crystals. The black line is sketch of the lower critical field $H_\mathrm{c1}$.The red circles are the $H_\mathrm{dip}$ obtained by the minimum value of $J_\mathrm{s}(H)$ at different temperatures. The green diamonds are the field $H_\mathrm{kink}$, which is the boundary of ordered-elastic region and disordered-elastic region. The blue uptriangles are the field $H_\mathrm{p}$, which is the boundary of elastic region and plastic region. The navy downtriangles are the irreversibility field $H_\mathrm{irr}$. The magenta squares are the upper critical field $H_\mathrm{c2}$.
} \label{fig5}
\end{figure*}

Fig.~\ref{fig5} shows the phase diagram of the Ba$_{0.64}$K$_{0.36}$Fe$_2$As$_2$ single crystal. The black line is the sketch of the lower critical field \(H_\mathrm{c1}(T)\), and the gray region below \(H_\mathrm{c1}(T)\) is the Meissner state. In the green region between \(H_{\text{kink}}(T)\) and \(H_\mathrm{c1}(T)\), when the magnetic flux lines initially penetrate into the bulk area of the sample, their motion is rapid, leading to the emergence of a peak at low field on the \(Q(H)\) curves \cite{ref10}. As the density of flux lines gradually increases, the vortex structure undergoes a phase transition from an ordered to a disordered phase, and a vortex glass state with vortex entanglement appears above \(H_{\text{kink}}\). In the blue region between \(H_{\text{kink}}(T)\) and \(H_\mathrm{p}(T)\), it is in the vortex glass state with strong entanglement of vortices. Above \(H_\mathrm{p}\), the entangled flux lines gradually break and cut each other, and the vortex system goes into a plastic flow regime. The irreversibility field \(H_{\mathrm{irr}}(T)\) is defined as the points where the zero-field-cooled (ZFC) and field-cooled (FC) magnetization intersect with each other, showing a boundary for the free flux flow. The upper critical field \(H_\mathrm{c2}(T)\) is defined as the starting point of the formation of the superconducting droplet, corresponding to the field where the magnetization deviates from the paramagnetic normal state background in \(M(T)\) curves. We define the yellow region between \(H_\mathrm{p}(T)\) and \(H_{\text{irr}}(T)\) as the plastic motion region and the magenta region between \(H_{\text{irr}}(T)\) and \(H_\mathrm{c2}(T)\) as the vortex liquid, respectively. Both regions are small. When the field exceeds \(H_\mathrm{c2}(T)\), the sample is in the normal state.
\section{Conclusions}
In summary, we have measured the dynamical and conventional magnetization relaxation rates of the Ba$_{0.64}$K$_{0.36}$Fe$_2$As$_2$ single crystal. The magnetic field dependence of both relaxation rates exhibits a minimum value at the same field. This point corresponds to a diminished vortex creep with a relaxation rate smaller than the so-called universal lower limit of the relaxation rate. This is attributed to the transition from ordered (Bragg glass) to disordered vortex (like vortex glass) phases. Above the peak position, the dislocations of the vortex proliferate quickly, leading to a rapid increase in the relaxation rate and a decrease in the critical current density. Further work is warranted, with a particular interest in the local magnetization measurements, which could more directly measure the order-disorder transition lines.
\begin{acknowledgments}
This work is supported by the National Natural Science Foundation of China (Grants No. A0402/11927809, No. A0402/11534005), National Key R and D Program of China (Grant No. 2022YFA1403201), and the Strategic Priority Research Program of Chinese Academy of Sciences (Grant No. XDB25000000).
\end{acknowledgments}    

$^\dag$ hhwen@nju.edu.cn

\end{document}